\documentclass[aps,prd,preprint,superscriptaddress,floatfix,nofootinbib]{revtex4}

\usepackage[dvips]{graphicx}

\newcommand{\lsim}{\mathrel{\mathop{\kern 0pt \rlap
  {\raise.2ex\hbox{$<$}}}
  \lower.9ex\hbox{\kern-.190em $\sim$}}}
\newcommand{\gsim}{\mathrel{\mathop{\kern 0pt \rlap
  {\raise.2ex\hbox{$>$}}}
  \lower.9ex\hbox{\kern-.190em $\sim$}}}

\def\ee{\end{equation}}
\def\be{\begin{equation}}
\def\eea{\end{eqnarray}}
\def\bea{\begin{eqnarray}}
\def\eeas{\end{eqnarray*}}
\def\beas{\begin{eqnarray*}}
\def\noi{\noindent}

\begin{document}

\preprint{DFTT 17/2002}
\preprint{INFNFE--06--02}

\title{~\\ Does Solar Physics Provide Constraints to Weakly
Interacting Massive Particles?}
 


%
\author{A. Bottino} 
\email{bottino@to.infn.it}
\affiliation{Dipartimento di Fisica Teorica, Universit\`a di Torino \\
Istituto Nazionale di Fisica Nucleare, Sezione di Torino \\
via P. Giuria 1, I--10125 Torino, Italy}

\author{G. Fiorentini} 
\email{fiorentini@fe.infn.it}
\affiliation{Dipartimento di Fisica, Universit\`a di Ferrara \\
Istituto Nazionale di Fisica Nucleare, Sezione di Ferrara \\
via del Paradiso 12, I--44100 Ferrara, Italy}

\author{N. Fornengo} 
\email{fornengo@to.infn.it}
\homepage{http://www.to.infn.it/~fornengo}
\affiliation{Dipartimento di Fisica Teorica, Universit\`a di Torino \\
Istituto Nazionale di Fisica Nucleare, Sezione di Torino \\
via P. Giuria 1, I--10125 Torino, Italy}

\author{B. Ricci} 
\email{ricci@fe.infn.it}
\affiliation{Dipartimento di Fisica, Universit\`a di Ferrara \\
Istituto Nazionale di Fisica Nucleare, Sezione di Ferrara \\
via del Paradiso 12, I--44100 Ferrara, Italy}

\author{S. Scopel} 
\email{scopel@to.infn.it}
\affiliation{Dipartimento di Fisica Teorica, Universit\`a di Torino \\
Istituto Nazionale di Fisica Nucleare, Sezione di Torino \\
via P. Giuria 1, I--10125 Torino, Italy}

\author{F.L. Villante} 
\email{villante@fe.infn.it}
\affiliation{Dipartimento di Fisica, Universit\`a di Ferrara \\
Istituto Nazionale di Fisica Nucleare, Sezione di Ferrara \\
via del Paradiso 12, I--44100 Ferrara, Italy}

\date{\today}

\begin{abstract} \vspace{1cm}  
  We investigate whether present data on helioseismology and solar
  neutrino fluxes may constrain WIMP--matter interactions in the range
  of WIMP parameters under current exploration in WIMP searches. We
  find that, for a WIMP mass of 30 GeV, once the effect of the
  presence of WIMPs in the Sun's interior is maximized, the squared
  isothermal sound speed is modified, with respect to the standard
  solar model, by at most 0.4\% at the Sun's center. The maximal
  effect on the $^8$B solar neutrino flux is a reduction of 4.5\%.
  Larger masses lead to smaller effects.  These results imply that
  present sensitivities in the measurements of solar properties,
  though greatly improved in recent years, do not provide information
  or constraints on WIMP properties of relevance for dark
  matter. Furthermore, we show that, when current bounds from direct
  WIMP searches are taken into account, the effect induced by WIMPs
  with dominant coherent interactions are drastically reduced as
  compared to the values quoted above. The case of neutralinos in the
  minimal supersymmetric standard model is also discussed.
\end{abstract}


\maketitle

\section{Introduction}
\label{sec:intro}

A host of independent astronomical observations point to the existence
in our Universe of a total amount of matter in the range $0.2 \lsim
\Omega_m \lsim 0.4$ \cite{matter} (or equivalently, $0.05 \lsim
\Omega_m h^2 \lsim 0.3$, where $h$ is the Hubble constant in units of
100 km $\cdot$ s$^{-1} \cdot$ Mpc$^{-1}$), well beyond the amount of
visible matter $\Omega_{vis} \sim 0.003$.  Since the primordial
nucleosynthesis tells us that baryons cannot contribute for more than
about 5\% \cite{bbn}, most of the dark matter must be
non-baryonic. The evolution theory of the primordial density
fluctuations into the present cosmological structures indicates that
most of the dark matter must be comprised of cold particles, {\it
i.e.} of particles that decoupled from the primordial plasma when
nonrelativistic. A particle with the suitable properties for being a
significant cold dark matter relic is generically defined as a WIMP
(weakly interacting massive particle).

A variety of  physical realizations for a WIMP are offered by various 
extensions of the Standard Model  \cite{nic}, the neutralino being
one of the most appealing candidates. In the present paper, many
considerations are developed in terms of generic WIMPs. We consider
the neutralino, whenever we wish to narrow down to a specific candidate.   

WIMPs are very actively searched for by means of various experimental 
strategies. Direct searches rely on the measurements of the signal
that a nucleus of an  appropriate detector  would generate, when hit by
a WIMP \cite{morales}. Indirect  searches are based on 
measurements of the signals due to WIMP pair annihilations in the galactic
halo or inside celestial bodies \cite{nic}. 

The event rates of WIMP direct search experiments are proportional to the
product of the local WIMP density $\rho_{\chi}$ in the Galaxy times
the WIMP-nucleus cross--section. In what follows $\rho_{\chi}$ will be
expressed as $\rho_{\chi} = \xi \cdot \rho_l$, where $\rho_l$ is the
local {\it total} dark matter density and $\xi$ ($\xi \leq 1$) is a
scaling parameter which accounts for the actual fraction of local
dark matter to be ascribed to the candidate $\chi$.

WIMP--matter interactions are conveniently classified in terms of
coherent cross--sections and (nuclear) spin--dependent ones.  In the
first case, the WIMP--nucleus cross--section is given in terms of the
WIMP--proton and WIMP--neutron cross--sections. To simplify the
formalism, in the following we further assume that, in the coherent
case, WIMPs interact with equal strength with protons and neutrons
(for instance, this may safely be assumed for neutralinos, while it is
not the case for neutrinos), and then we express a generic
WIMP--nucleus coherent cross--sections in terms of a single
WIMP--nucleon cross--section $\sigma_c$.  For the spin--dependent
case, the derivation of a WIMP--nucleon cross--section from the
WIMP--nuclear one depends on the nature of the WIMP and on specific
nuclear properties \cite{bdmsbi}.

In the case of WIMPs with coherent interactions with matter, the range
of WIMP parameters under current exploration in WIMP direct searches
is conveniently expressed in terms of the quantity $\xi \sigma_c$:
\begin{equation}
10^{-43} \; {\rm cm^2} \leq \
\xi \sigma_c \leq 
 6 \cdot 10^{-41} \; {\rm cm^2} \, ,
\label{eq:sens}
\end{equation}
for WIMP masses $m_{\chi}$ in the range:
\begin{equation}
30 \; {\rm GeV} \leq  m_\chi \leq 270 \;  {\rm GeV}. 
\label{eq:mass}
\end{equation}      

In the derivation of the ranges in
Eqs. (\ref{eq:sens}--\ref{eq:mass}), one has taken into account a
variety of WIMP distribution functions in the galactic halo and
uncertainties in the determination of the relevant astrophysical
quantities \cite{belli1,belli2}.

For the reasons mentioned above, in the case of spin--dependent
interactions no model--independent sensitivity range may be derived.
However, to give an indication, we recall that, in the case of a relic
particle which interacts with matter mainly by spin-dependent
interactions mediated by $Z$--boson exchange, current WIMP direct
experiments are sensitive to values of $\xi \sigma_{sd}$ of about
$10^{-36} - 10^{-37}$ cm$^2$ \cite{bdmsbi}, where $\sigma_{sd}$ is the
spin--dependent WIMP-proton cross--section.

One of the WIMP direct search experiments, the DAMA/NaI(Tl) experiment
\cite{dama}, has observed an annual modulation effect (at a $4\sigma$
C.L.) with all the features expected for a WIMP signal
\cite{freese}. When interpreted as due to a WIMP with coherent
interactions, the DAMA effect provides a $3\sigma$ region in the plane
$m_{\chi} - \xi \sigma_c$ embedded in the range of
Eqs.(\ref{eq:sens}--\ref{eq:mass}).  In Ref. \cite{noi} it was proved
that this annual modulation effect is compatible with an
interpretation in terms of relic neutralinos (see also
Ref.\cite{others}).
 
If WIMPs populate our Galaxy, they can be captured by the Sun and
accumulate in its core. For suitable values of the WIMP parameters,
this could have relevant effects on the Sun. In fact, WIMPs could
provide an effective mechanism for energy transport in the Sun,
producing an isothermal core and reducing substantially the Sun
central temperature $T_c$.  Following this idea, some time ago a
special class of WIMPs named {\it cosmions}, with masses of a few GeV
and scattering cross sections on nucleons of the order of $10^{-36}$
-- $10^{-34}$ cm$^2$, was studied in detail as a way of solving
simultaneously the solar neutrino puzzle and the dark matter problem
(see e.g. Refs.
\cite{Gould:hm,Spergel:1985,Gould:1989ez,Gilliland:1986,Dearborn:1990mm,cosmions_others}). The
cosmion hypothesis was progressively abandoned when it became clear
that the solar neutrino puzzle cannot be accounted for by simply
reducing $T_{c}$.

In the last twenty years, our observational knowledge of the solar
interior has progressed enormously. By means of helioseismic data it
has become possible to derive the sound speed with an accuracy of
about one part per--thousand over most of the radial profile and of
about one percent in the innermost part \cite{Ricci:1997}. By the same
method, it has been possible to deduce important properties of the
convective envelope.  The photospheric helium fraction $Y_{ph}$ and
the depth of the convective envelope $R_{b}$ have been determined with
an accuracy of about one per cent and one per--thousand respectively,
following the pioneering papers of
Refs. \cite{Dziembowski:1991,Christensen:1991}.  Moreover, the
measurement of the neutrino flux from $^8$B decay obtained by
combining
\cite{Villante:1998pe,Fiorentini:2001jt,Fogli:2001vr,Fogli:2002ts} SNO
charged current \cite{Ahmad:2001an} and Super-Kamiokande
\cite{Fukuda:2001nj} data and confirmed by the recent SNO neutral
current results \cite{Ahmad:2002jz}, has provided a determination of
the temperature $T_{c}$ near the center of the Sun at the level of
about one per cent \cite{Fiorentini:2001et}.

All the predictions of the Standard Solar Model (SSM) have been
confirmed by these accurate tests, so one is naturally led to the
question of whether our accurate knowledge of the Sun interior can be
used to constrain the WIMP parameter space.  This particularly
interesting question was recently raised in Refs. \cite{Lopes:2001ra}
and \cite{Lopes:2001ig}, where it was concluded that solar physics can
be used to significantly constrain the WIMP parameter space.

In this paper, we provide an alternative analysis of the
problem, obtaining results substantially different from those derived
in Refs. \cite{Lopes:2001ra,Lopes:2001ig}.

The plan of the paper is as follows. In Section II we discuss WIMP
energy transport in the Sun, introducing the relevant physical
parameters to be discussed in the subsequent sections. In Section III
and IV WIMP energy transport is applied to the central solar
structure, in order to determine the regions of the WIMP parameter
space where the WIMP energy transport is most efficient. This leads to
an optimal choice of WIMP parameters which is then used in Section V
to calculate the solar sound speed profile and discuss the possible
impact of helioseismology on WIMP properties. The sensitivity of the
available data on $^8$B solar neutrinos to the presence of WIMPs in
the Sun is then discussed in Section VI. The neutralino, as a specific
realization of WIMP, is discussed in Section VII. Section VIII is
devoted to our conclusions
 
\section{WIMP energy transport}
\label{sec:transport}

In this section, we briefly review some issues relevant to the problem
of WIMP energy transport in the Sun addressed in
Refs. \cite{Gould:hm,Spergel:1985,Gould:1989ez,Gilliland:1986,Dearborn:1990mm,cosmions_others}.

WIMPs are captured by the Sun during the course of its lifetime and
are confined by the effect of gravity in a central region with a
radial number density distribution approximately given by
\cite{gr_seck,press_spergel_2} (we hereby use natural units:
$\hbar=c=k_B=1$):
\be
n_{\chi}(r) = (n_{\chi})_c \; \exp\left(-\frac{r^2}{r_{\chi}^2}\right),
\label{radial}
\ee 
where the quantity $(n_{\chi})_c = (1/\pi^{3/2})\; (N_{\chi}/r_{\chi}^3)$
 is the
central WIMP number density, $N_{\chi}$ is the total number of WIMPs in
the Sun and  
\be r_{\chi}
= \left(\frac{3 T_{c}}{ 2\pi G \rho_{c} m_{\chi}}\right)^{1/2},
\label{r_x}
\ee 
$T_{c}$ and $\rho_c$ being the Sun central temperature and density,
and $m_\chi$ the WIMP mass. For $T_c = 1.57\cdot 10^{7}$ K and
$\rho_c=154$ g cm$^{-3}$, one finds $r_{\chi}\simeq 0.01\, ({\rm 100\;
GeV} / m_{\chi})^{1/2} R_{\odot}$.  We note that Eq.~(\ref{radial}) is
obtained under the assumption of constant density and temperature in
the region of the Sun populated by WIMPs. This approximation is
accurate enough for our purposes.

WIMPs provide a mechanism for energy transport in the solar
core whose efficiency depends on the Knudsen parameter
\be
K = \frac{l_{\chi}(0)}{r_{\chi}}~,
\ee
where $l_{\chi}(0)$ is the WIMP mean free path near the solar center
and generally:
\be
\frac{1}{l_{\chi}(r)} = \sum_{i} \sigma_{i} X_{i}(r) \,\frac{\rho(r)}{m_i}.
\label{eq:mean_free_path}\ee
Here the sum extends to all the nuclear species present in the Sun,
$X_{i}$ is the mass fraction of the $i$--th element, $\rho(r)$ is the
density profile of the Sun and $\sigma_{i}$ is the WIMP scattering
cross section with the nucleus of species $i$.

For $K \ll 1$ the WIMP mean free path is much smaller than the
dimension of the region where the WIMPs are confined and the energy
transfer mechanism is {\it conductive}. The problem of thermal
conduction by a dilute gas of massive particles is discussed in Ref.
\cite{Gould:hm}.  For our purposes, it is useful to express the WIMP
luminosity in the conductive regime by using the radiative transport
equation: 
\be L_{\rm \chi,cond}(r) = - \frac{16\pi a_{\rm rad}}{3} \;
\frac{T(r)^3 r^2}{\kappa_{\chi}(r) \, \rho(r)} \;\frac{dT}{dr},
\label{lcond}
\ee
where the WIMP opacity $\kappa_{\chi}$ is defined as:
\be \frac{1}{\kappa_{\chi}(r)} = \frac{3 n_{\chi}(r)}{4 a_{\rm rad}
T(r)^3} \, \sqrt{\frac{T(r)}{m_{\chi}}} \, \rho(r) \, l_{\chi}(r) \,
\lambda_{\chi} \, .
\label{kappax}
\ee
In Eqs. (\ref{lcond},\ref{kappax}) the quantity $a_{\rm rad}$ denotes
the radiation density constant and $\lambda_{\chi}$ is a dimensionless
coefficient which depends on the ratio between the WIMP mass and the
masses of background nuclei
\footnote{The coefficient $\lambda_{\chi}$ can be derived from the
values of the thermal conduction coefficient $\kappa(m_{\chi}/m_i)$
tabulated in Ref. \cite{Gould:hm}. It holds: $\lambda_{\chi} =
(1/l_{\chi})[\sum_{i} (1/l_{\chi,i}\,\kappa)]^{-1}$ where $l_{\chi,i}$
is the WIMP mean free path relative to the nucleus $i$.}.

For $K \gg 1$ the WIMP mean free path is much larger than the region
in which WIMPs are trapped and the energy transfer mechanism is {\it
non-local}.  An analytic approximation to treat this case has been
developed by Spergel and Press \cite{Spergel:1985} who described WIMPs
as an isothermal gas at an appropriate temperature $T_{\chi}$. In this
case, the luminosity $L_{\rm \chi, NL}$ carried by WIMPs can be
expressed as: 
\be L_{\rm \chi,NL}(r) = \int^{r}_{0} dr' \; 4 \pi
r'\,^{2} \, \rho(r') \, \epsilon_{\rm SP}(r';T_{\chi}),
\label{lnl}
\ee
where $\epsilon_{\rm SP}$ represents the energy transferred to WIMPs
per second and per gram of nuclear matter and is given by:
\begin{eqnarray}
\nonumber
\epsilon_{\rm SP}(r;T_{\chi}) & = & 
8~\sqrt{\frac{2}{\pi}} \;
n_{\chi}(r) \times \\
& & \sum_{i} \sigma_{i}   \frac{X_{i}(r)}{m_{i}} \,
\frac{m_{\chi} m_i}{(m_{\chi} + m_i)^2}
\left( \frac{m_{\chi}T(r) + m_{i}T_{\chi}}{m_{\chi} m_{i}} \right)^{1/2}
\left[ T(r) - T_{\chi} \right] ~.
\label{eps_sp}
\end{eqnarray}
The WIMP temperature $T_{\chi}$ is fixed by requiring
that the energy absorbed by WIMPs in the inner region
(where $T(r) > T_{\chi}$) is equal to that released in the outer region
(where $T(r) < T_{\chi}$), {\it i.e.}:
\be
L_{\rm \chi,NL}(R_{\odot}) = 
\int^{R_{\odot}}_{0} dr' \; 4 \pi r'\,^{2} \rho(r') \,
\epsilon_{\rm SP}(r';T_{\chi})  = 0 ~.
\ee

Gould and Raffelt \cite{Gould:1989ez}, by means of a Montecarlo
integration of the Boltzmann collision equation, showed that the
Spergel and Press approximation, Eq.(\ref{eps_sp}), overestimates the
energy transfer typically by a factor of a few.  Moreover, by studying
simplified stellar models, they showed that it is possible to
approximate energy transfer in the non-local regime by a conductive
treatment if one applies a global ``luminosity suppression
factor''. We follow their approach by using as a general expression
for the WIMP luminosity:
\be
L_{\chi}(r) = f(K) \; L_{\rm \chi, cond}(r) \, ,
\label{general}
\ee 
where 
\be 
f(K) = \frac{1}{(K/K_{0})^{2}+1} \, ,
\ee 
and $K_0 = 0.4$ is the value of the Knudsen parameter for which
the WIMP energy transport is most efficient.  We remark that other
approaches are possible. In Refs. \cite{Gilliland:1986} and
\cite{Lopes:2001ra,Lopes:2001ig}, for example, a suitable interpolation
of Eqs.~(\ref{lcond}) and (\ref{lnl}) is used as a general expression
for WIMP luminosity.  We have checked numerically that our results are
essentially independent of the chosen approach.

As mentioned before, WIMPs can interact with nuclei either via
coherent interactions or via spin--dependent ones. In the first case
the WIMP cross section on the $i$--th nucleus of mass $m_i$ is related
to the corresponding WIMP-nucleon cross--section, $\sigma_c$, by:
\be 
\sigma_i= \sigma_{c}
\left(\frac{m_{i}}{m_n}\right)^{4}\left(\frac{m_{\chi}+m_n}{m_{\chi}+
m_{i}}\right)^{2}~,
\label{eq:sigmac}
\ee 
where $m_n$ is the nucleon mass.
With this assumption, Eq. (\ref{eq:mean_free_path})
may be rewritten as:
\be
\frac{1}{l_{\chi}(r)} = \frac{1}{l_n (r)} \sum_{i} X_{i}(r) 
\left ( \frac{m_i}{m_n}\right )^3
\left ( \frac{m_{\chi}+m_n}{m_{\chi}+m_i}\right )^2 \, ,
\label{eq:mean_free_path2}
\ee
with:
\be
 l_n(r)=\frac{m_n}{\sigma_c ~\rho(r)} \simeq \left ( \frac{\rm 1.6 \cdot
10^{-37} \;cm^2}{\sigma_c} \right ) \cdot \left ( \frac{\rm 150
\;g\;cm^{-3}}{\rho(r)}\right) \cdot R_{\odot}. 
\ee
In the Sun the abundances $X_i$ for nuclei heavier than Hydrogen and
helium are much smaller than those of H and He. However, in the sum
over elements of Eq.(\ref{eq:mean_free_path2}), for heavy nuclei
(mainly for iron) the reduction due to $X_i$ is largely compensated by
the coherence effect displayed by Eq.(\ref{eq:sigmac}).

On the other hand, as far as spin--dependent interactions are
concerned, the contribution of Hydrogen is overwhelmingly dominant
over the contributions of other nuclear species with non--vanishing
spins. Therefore, from now on we identify $\sigma_{sd}$ with the
spin--dependent WIMP-proton cross-section.

In the following all WIMP matter interactions will be expressed in
terms of $\sigma_c$ and/or $\sigma_{sd}$. The notation $\sigma_p$ will
be used to denote a generic WIMP--proton cross-section; $\sigma_p$ has
to be identified with $\sigma_c$ (or $\sigma_{sd}$), whenever coherent
(or spin--dependent) interactions are considered.

\section{WIMPS and the central solar structure}
\label{sec:structure}

The efficiency of WIMP energy transport depends on several
parameters. Specifically, it depends on the total number $N_{\chi}$ of
WIMPs in the Sun, on the WIMP mass $m_{\chi}$ and on the WIMP-nucleon
cross section $\sigma_p$. In order to determine the region of the
WIMP parameter space where the WIMP energy transport is most efficient
it is useful to define the quantity:
\be
\delta = \lim_{r\rightarrow 0}  \frac{L_{\chi}(r)}{L(r)} \, , 
\label{eq:delta}
\ee
where $L$ is the radiative luminosity of the Sun.
It is easy to show that:
\be
\delta = \left[\frac{1}{(K/K_{0})^{2}+1} \right]\;
\frac{\kappa_{\gamma}(0)}{\kappa_{\chi}(0)} \, ,
\label{delta} 
\ee
where 
$\kappa_{\gamma,\chi}(0)$ are the opacities at the solar center for
radiation and WIMPs.

In Figs. \ref{fig:coer_m50}--\ref{fig:spin_m50} we show the lines in
the plane $(\sigma_p,N_{\chi})$ which correspond to fixed values of
$\delta$.  We have chosen $m_{\chi}=50$ GeV as a representative value
for the WIMP mass.  Fig. \ref{fig:coer_m50} refers to coherent WIMP
scattering, while Fig. \ref{fig:spin_m50} is obtained for
spin--dependent WIMP interactions. The lines have been calculated by
using the physical and chemical parameters of the Sun as they are
obtained by a standard evolutionary code \cite{Ciacio:1996mk}.

The qualitative behavior of the iso-$\delta$ lines is easily
understood. The efficiency of the WIMP energy transport is
proportional to $1/\sigma_p$ ({\it i.e.} to $K$) in the conductive
regime and to $\sigma_p$ ({\it i.e.} to $K^{-1}$) in the non-local
regime.  The transition between the two regimes occurs at $K =
l_{\chi} / r_{\chi} \simeq 0.4$, {\it i.e.} at a cross--section
$\sigma_t$ which corresponds, for coherent scattering, to the value
$\sigma_{t} \simeq 10^{-37}$~cm$^{2}$ and, for spin--dependent
interactions, to $\sigma_{t} \simeq 10^{-34}$~cm$^{2}$. The difference
in the values of the transition cross sections $\sigma_{t}$ is due to
the fact that, for the spin--independent case, interactions on
elements heavier than Hydrogen is important, due to the coherence
effect in the scattering which is manifest in Eq. (\ref{eq:sigmac}),
while for spin--dependent interactions no coherence is present and
therefore only scattering on Hydrogen (which is largely more abundant
than other elements with non vanishing spin) matters.

To be more quantitative, we can also show the analytic expression of
the iso-$\delta$ lines as a function of $\sigma_p$:
\be N_{\chi}(\delta,\sigma_p) = N(\delta)
\left[(\sigma_p/\sigma_t)+(\sigma_p/\sigma_t )^{-1}\right],
\label{eq:sigma_t}
\ee
where $N(\delta)$ is given by the condition:
%
%
\be
\kappa_{\gamma}(0) = \delta \; \kappa_{\chi}(0; l_{\chi} = 0.4\,r_{\chi}) \, ,
\ee
which gives:
\be \frac{N(\delta)}{\pi^{3/2} \, r_{\chi}^3} = \delta \, \frac{4
a_{\rm rad} T_c^3}{3} \, \sqrt{\frac{m_{\chi}}{T_c}}\,
\frac{l_{\gamma}}{0.4\,r_{\chi}\,\lambda_{\chi}} \, , \ee
and $l_{\gamma}\simeq 5\cdot 10^{-3}~{\rm cm}$ is the photon mean free
path at the center of the Sun.

We remark that, since WIMP radial distribution is rapidly decreasing 
(see Eq.(\ref{radial})), the parameter $\delta$ essentially
provides an upper limit 
for the ratio $L_{\chi}/L$ in the Sun. 
As a consequence, the condition that $\delta$ is not
too small should be used as a {\it necessary (but not sufficient)}
condition for WIMPs to have relevant effects on the Sun. 
This means that:

{\it i)} {\it We can safely conclude that regions of the parameter
space for which $\delta \lsim 10^{-2}$ are not interesting.}  This
point can be easily understood. WIMPs modify the relation between the
luminosity and the gradient of temperature in the Sun. This
essentially corresponds to a redefinition of the Sun radiative opacity
$\kappa_{\gamma}$.  The effect is maximal at the center of the Sun
where one has $\kappa_{\gamma} \rightarrow
\kappa_{\gamma}/(1+\delta)$.  The radiative opacity is affected by
uncertainties at the level of a few percent \cite{opacity}.  This
means that we cannot disentangle WIMP effects if $\delta$ is smaller
than $10^{-2}$.

{\it ii)} {\it We cannot conclude, at this stage, that regions in
which $\delta$ is large (say e.g. $\delta > 1$) correspond to sizeable
modifications of the solar structure.}  The parameter $\delta$ is, in
fact, a local parameter. A large $\delta$ only implies that WIMPs give
a large contribution to the energy transport at the center of the
Sun. In order to understand whether WIMPs have relevant effects on
global properties of the Sun, we also have to compare the radius
$r_\chi$ of the WIMPs extension region with the relevant length scales
in the Sun ({\em e.g.} the dimension of the neutrino production
region, the temperature height scale).

\section{The number of WIMPs in the Sun}
\label{sec:number}

Up to this point we have treated $m_{\chi}$, $\sigma_p$ and $N_{\chi}$
as independent quantities.  However, this is not the case, as noted
long ago in Ref. \cite{Dearborn:1990mm}.  $N_{\chi}$ depends on the
capture rate of the halo WIMPs by the Sun during its lifetime; in
turn, this capture rate is proportional to the product of the WIMP
density in the halo $\rho_{\chi}$ times $\sigma_p$, {\it i.e.} to the
product $\xi \sigma_p$. Moreover, captured WIMP may annihilate with
captured antiWIMP. Therefore $N_{\chi}$ also depends on the quantity
$\langle \sigma_a v \rangle_0$, where $\sigma_a$ is the WIMP-antiWIMP
pair annihilation cross--section and $v$ is the relative velocity of
the annihilating pair, while brackets indicate average over the WIMP
velocity distribution in the Sun core. Explicitly, one has
\cite{gr_seck}:
\be 
N_{\chi} = \Gamma_c \, \tau \, \tanh(t_{\odot}/\tau), 
\label{eq:n_chi}
\ee 
where $t_{\odot} \simeq 4.5$ Gyr is the age of the Sun, $\Gamma_c$ is
the WIMP capture rate and $\tau=1/\sqrt{\Gamma_c C_a}$ is a time scale
parameter which determines whether and when equilibrium between
capture and annihilation occurs. The quantity $C_a$ is given by:
\be C_a=\frac{\langle \sigma_a v \rangle_0}{V_0} \left (
 \frac{m_{\chi}}{{\rm 20\; GeV}}\right)^{3/2}, \ee
\noi with $V_0\simeq 2.3 \times 10^{28}$ cm$^3$ for the Sun.  In the
limit in which $ \langle \sigma_{a} v \rangle_0 \rightarrow 0$, one has
$\tau \rightarrow \infty$ and Eq. (\ref{eq:n_chi}) becomes:
\be N_{\chi} = \Gamma_c \, t_{\odot} \equiv N_\chi^{\rm max} .
\label{t_evol}
\ee
$N_\chi^{\rm max}$ represents the upper limit for the WIMP number in the
Sun at present time.

The expression of the capture rate $\Gamma_c$ is rather involved. For
a Maxwellian WIMP velocity distribution in the galactic halo it may
be cast in the form \cite{Gould:1987}:
\be \Gamma_c = \sum_i \left ( \frac{8}{3 \pi}\right )^{1/2} \left [
\sigma_i \frac{\rho_{\chi}}{m_{\chi}} \bar{v} \right ] \left
[\frac{M_i}{m_i}\right ] \left [ \frac{3 v_{esc}^2}{2 \bar{v}^2}
\langle \phi \rangle_i \right ] \xi(\infty) S_i \;\;\; ,
\label{eq:capture}
\ee
where the cross sections $\sigma_i$ are related to $\sigma_p$ as
discussed in the previous Section, $M_i$ is the total mass of the
$i$--th nucleus in the Sun, $v_{esc}\simeq$ 618 km sec$^{-1}$ is the
escape velocity at the Sun's surface and $\bar{v} = 270$ Km s$^{-1}$
is the WIMP velocity dispersion. The quantities $\langle \phi
\rangle_i$ denote the reduced gravitational potential $\phi(r) =
v_{esc}^2(r)/v_{esc}^2$ averaged over the mass distribution of the
$i$--th nucleus, as defined in Eq. (\ref{eq:mass_ave}) in the
Appendix. The quantity $\xi(\infty) \simeq$ 0.75 is a suppression
factor due to the motion of the Sun through the halo and the factor
$S_i$ includes the effects of WIMP--nucleus mass mismatch and finite
target dimensions (note that the $\sigma_i$ are point--like
cross--sections). The expression of $S_i$ is given in the Appendix for
completeness. The chemical composition of the Sun and the values of
the quantities $\langle \phi \rangle_i$, which have been used in our
calculations, are given in Table \ref{tab:sun}. We note that for
relatively heavy WIMPs like the ones which are considered in this
analysis ($m_\chi \gsim 30$ GeV) the capture in the Sun for
spin--independent interaction is not dominated by scattering on
Hydrogen, even though this is the most abundant element, but instead
by interactions on heavier elements. For instance, for a 50 GeV WIMP,
capture is dominated by He, O and Fe. The smaller abundance of heavier
elements is largely compensated by the coherence effect in the cross
section shown in Eq. (\ref{eq:sigmac}). On the contrary, in the case
of spin--dependent interactions the only relevant capture process is
on Hydrogen, since in this case no coherence effect is present and
therefore capture is dominated by interactions with the most abundant
element which possesses spin, {\em i.e.} Hydrogen.

Eq. (\ref{eq:capture}) holds when the cross section $\sigma_p$ is
small enough so that multiple scatterings can be neglected.  However,
when $\sigma_p$ is so large that every WIMP crossing the Sun is
captured, $\Gamma_c$ saturates to a maximal value, and
Eq. (\ref{eq:capture}) is replaced by:
\be
\Gamma_{c,sat} = \left ( \frac{8}{3 \pi}\right )^{1/2}
\left [ \frac{\rho_{\chi}}{m_{\chi}} \bar{v} \right ]
\left [\frac{M_i}{m_i}\right ]
\left [\zeta + \frac{3 v_{esc}^2}{2 \bar{v}^2} \right ]
\xi(\infty) \pi R^2_{\odot} \;\;.
\label{eq:saturation}
\ee
where $\zeta \simeq 1.77$. In Eq. (\ref{eq:saturation}), the first
term in the last square brackets refers to WIMPs whose orbit cross the
Sun even without gravitational deflection, while the second term is a
gravitational focusing factor and corresponds to those WIMPs whose
orbits pass through the Sun because of gravitational deflection. One
can easily see that for the Sun the focusing factor is dominant over
the purely ``geometrical'' one, since the velocity dispersion of WIMPs
in the Galaxy is a factor of two smaller than the escape velocity at
the Sun's surface.

\begin{figure} \centering
  \includegraphics[width=1.1\textwidth]{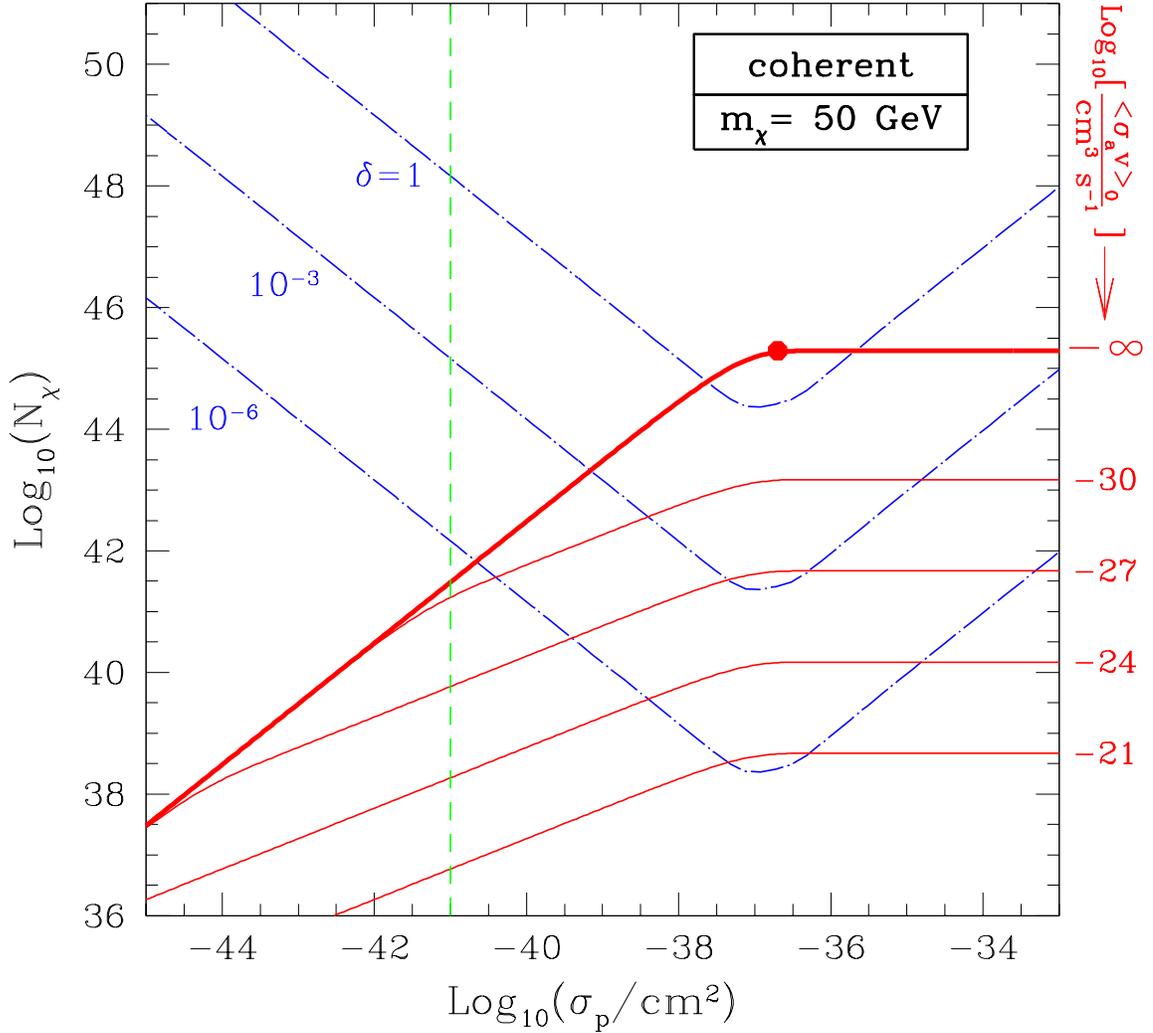} 
\vspace{-55pt}
\caption{The solid (red) curves denote the total number $N_\chi$ of
WIMPs accumulated in the Sun's core due to gravitational capture as a
function of the WIMP--nucleon cross section, for a WIMP mass of 50 GeV
and for the case of purely coherent WIMP--nucleus interactions. The
thick (red) solid line refers to WIMPs which do not annihilate once
captured. The thin (red) solid lines refer to annihilating WIMPs, for
increasing values of the zero--temperature thermally--averaged
WIMP--antiWIMP annihilation cross section: $\langle \sigma_a v
\rangle_0 = 10^{-30}$, $10^{-27}$, $10^{-24}$, $10^{-21}$ cm$^{3}$ s
$^{-1}$. The dot--dashed (blue) lines show the contours of constant
values of the parameter $\delta$, defined in Eq.(\ref{eq:delta}),
which quantifies the efficiency of energy transport of WIMPs with
respect to photons at the center of the Sun. The vertical (green)
dashed line denotes the upper limit on the WIMP--nucleon elastic cross
section for coherent interactions. The large (red) dot identifies the
WIMP parameters for which the WIMP effect on the Sun's energetic is
maximal.
\label{fig:coer_m50}}
\end{figure}

\begin{figure} \centering
  \includegraphics[width=1.1\textwidth]{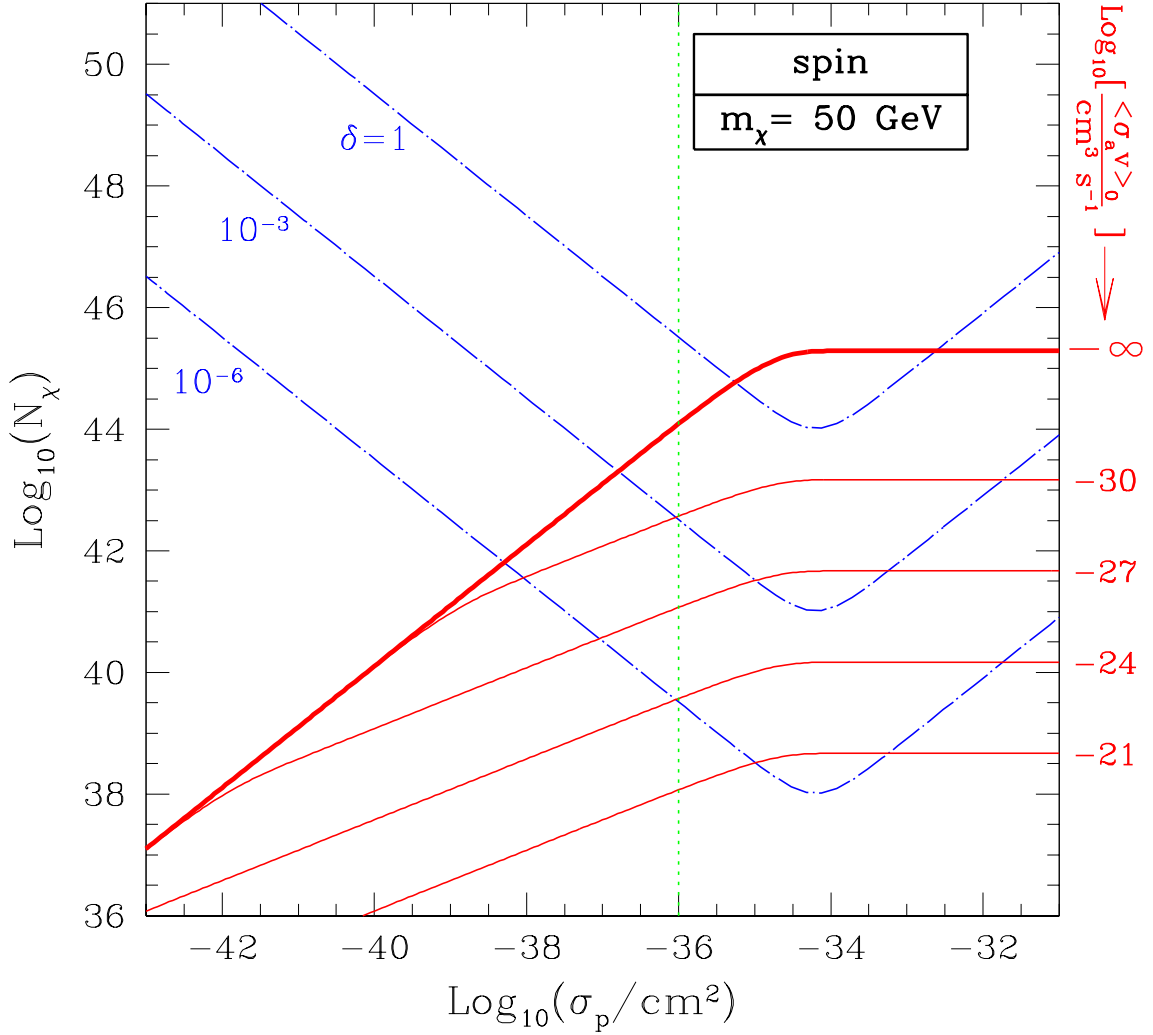}
\vspace{-50pt} 
\caption{The same as in Fig. \ref{fig:coer_m50}, for spin--dependent
WIMP--nucleus interactions. The vertical (green) dotted line denotes
the upper limit on the WIMP--nucleon cross section for $Z$--boson
mediated spin--dependent interactions.
\label{fig:spin_m50}}
\end{figure}

The behaviour of the capture rate, and consequently of the total
number $N_\chi$ of WIMPs captured by the Sun, as a function of the
WIMP parameters is now easily understood. At fixed $m_{\chi}$,
$N_{\chi}$ increases linearly with $\sigma_p$ up to the saturation
level. For $m_\chi = 50$ GeV, this behaviour can be seen in
Fig. \ref{fig:coer_m50} (coherent interactions) and
Fig. \ref{fig:spin_m50} (spin--dependent interactions), where the
thick solid line represents $N_{\chi}$ as a function of $\sigma_p$ for
the case of no--annihilation ($\sigma_a = 0$) and the thin solid lines
show $N_{\chi}$ for some representative non--vanishing values of
$\langle \sigma_a v \rangle_0$. In calculating $N_{\chi}$ as a
function of $\sigma_p$, we have set $\rho_{\chi}=\rho_l$, {\it i.e.}
$\xi=1$.  For $\rho_l$ we have used, here and throughout the paper,
the default value $\rho_l=0.3$ GeV cm$^{-3}$. The current uncertainty
on $\rho_l$ (0.2 GeV cm$^{-3}$ $\lsim \rho_l \lsim$ 0.7 GeV cm$^{-3}$,
for an isothermal galactic halo \cite{belli2}) may actually increase
or reduce $N_\chi$ by at most a factor of 2. The no--annihilation line
represents, for a given WIMP mass, the maximal number of WIMPs that
can be captured by the Sun during its lifetime. If annihilation is
present, $N_{\chi}$ is obviously smaller, depending on the strength of
the annihilation cross section. Results very similar to those shown in
Figs. \ref{fig:coer_m50}, \ref{fig:spin_m50} are obtained for
different WIMP masses. Notice that the transition from the linear
regime to the saturation level occurs at quite different values for
the relevant WIMP--nucleon cross section in the two limiting cases of
pure coherent and pure spin--dependent interactions.  This is again
due to the fact that for WIMPs heavier than tens of GeV, scattering on
heavy nuclei in the Sun is enhanced for coherent interactions, and
this shifts the transition from the linear to the saturation regime
toward WIMP--nucleon cross sections smaller than in the
spin--dependent case.

\begin{figure} \centering
  \resizebox{1.0\textwidth}{!}{\includegraphics[130,250][545,642]{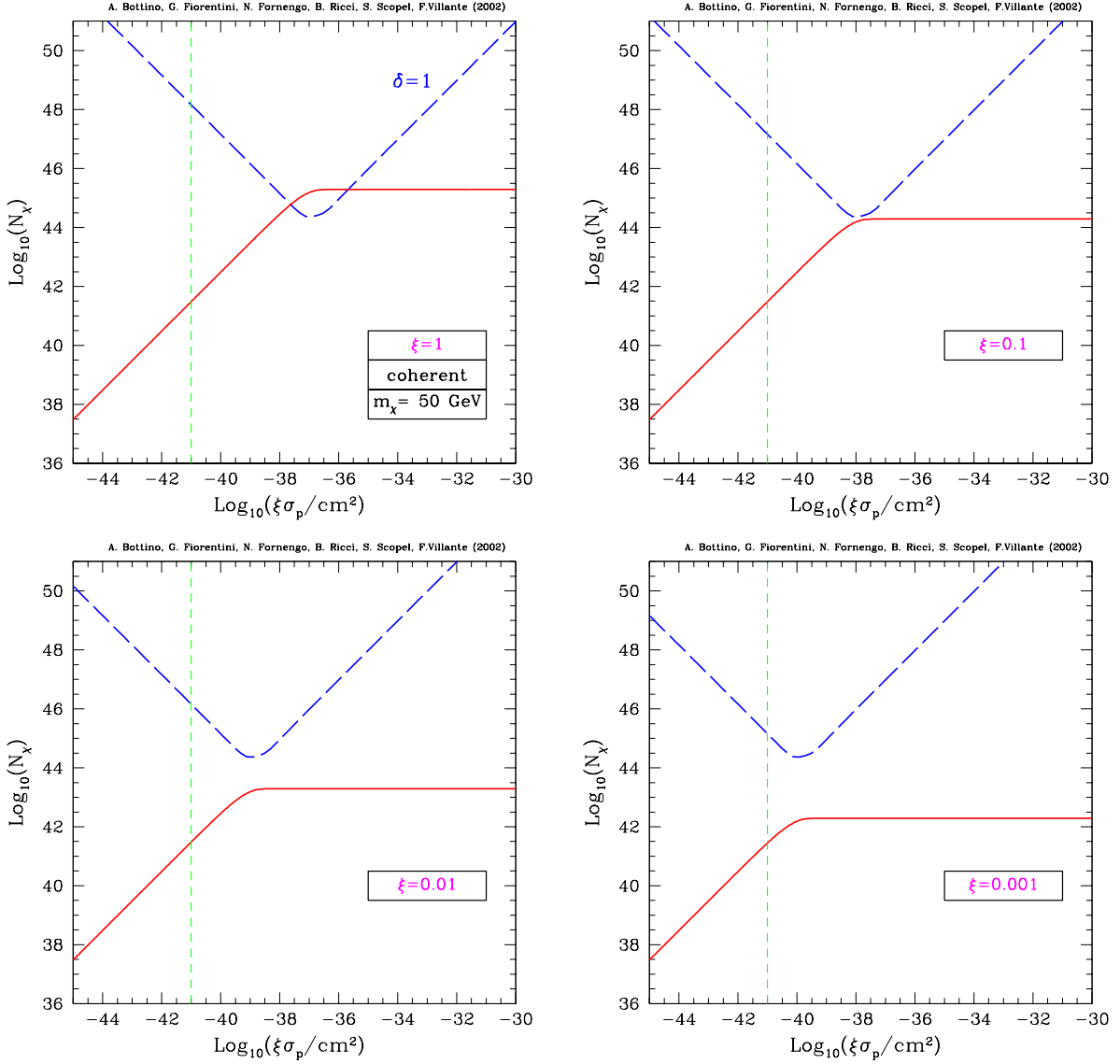}}
\caption{The same as in Fig. \ref{fig:coer_m50}, plotted as a function
of $\xi \sigma_p$, for different values of the rescaling parameter
$\xi$. The (red) solid lines denote the total number of WIMPs
accumulated in the Sun for a non--annihilating WIMP. The (blue) dashed
lines show the contours of iso--$\delta$ for $\delta=1$. The vertical
(green) dashed lines indicate the upper limit on $\xi \sigma_p$ for
coherent interactions.
\label{fig:coer_m50_rescal}}
\end{figure}

The previous discussion may be easily extended to the case in which
the WIMP under consideration does not provide the total amount of
local dark matter ({\it i.e.} $\xi<1$). In this case, the relevant
quantities are conveniently plotted in the plane $\xi
\sigma_p$--$N_{\chi}$. For the case of coherent interactions the
results are displayed in Fig. \ref{fig:coer_m50_rescal}, whose four
panels refer to different representative values: $\xi=1$, 0.1, 0.01,
0.001. In each panel the dashed curve denotes a representative
iso--$\delta$ line with $\delta=1$ and the solid line displays
$N_{\chi}$ as a function of $\xi \sigma_p$ for the limiting case
$\sigma_a=0$. For the sake of comparison, the top--left panel repeats
the case already shown in Fig. \ref{fig:coer_m50} ($\xi=1$). The
features of the panels with $\xi<1$ are easily understood in terms of
those represented in the top--left panel: i) the iso--$\delta$ curve
simply shifts horizontally to the left by an amount equal to $\xi$,
since $\delta$ depends on the scattering cross section $\sigma_p$ but
it does not depend on $\xi$; ii) the slanted part of the solid line is
unchanged since $\Gamma_c$ in the linear regime, as given by
Eq.(\ref{eq:capture}), is a function of the product $\xi \sigma_p$;
iii) the flat part of the solid line (saturation regime) is lowered by
an amount $\xi$ since $\Gamma_{c,sat}$ in Eq.(\ref{eq:saturation})
depends on $\xi$ but is independent of $\sigma_p$. We explicitely
consider the possibility of subdominant WIMPs since this is a feature
which can naturally occur for specific dark matter candidates, like,
for instance, the neutralino, as it will be discussed in
Sect. \ref{sec:neutralino}.

We are now able to understand which regions of WIMP parameter space
potentially lead to observable modification of the solar structure, by
comparing the iso--$\delta$ lines with the curves which describe the
number of WIMPs in the Sun. The first result is that the WIMP maximum
effect occurs when annihilation is negligible and for scattering cross
sections in the interval $10^{-38}~{\rm cm^2} \lsim \sigma_c \lsim
10^{-36}~{\rm cm^2}$ for coherent interactions and $10^{-35}~{\rm
cm^2} \lsim \sigma_{sd} \lsim 10^{-33}~{\rm cm^2}$ for spin--dependent
interactions.  These cross sections correspond to a situation in which
the number of WIMPs in the Sun has essentially reached its saturation
value and the WIMP energy transport is maximally efficient ({\em i.e.}
$K\sim K_{0}$).  This is manifest in Figs. \ref{fig:coer_m50},
\ref{fig:spin_m50} for $m_\chi=50$ GeV and occurs also for all the
WIMP mass range $m_\chi \gsim 30$ GeV under study in the present
paper. The value of $\delta$ which corresponds to this maximal effect
is $\delta\sim10$, both for coherent and spin--dependent
interactions. We will use this information in the next Sections where
we will calculate the {\em maximal} deviation of the Sun properties
when WIMPs are incorporated in its interior.

We stress that, for the WIMP mass range considered in the present
paper, the process of WIMP capture cannot produce modification of the
Sun which refer to values of $\delta$ larger than about 10, for any
value of the scattering cross--section. Independent limits on the
WIMP--nucleon interactions, like the ones that are obtained from
direct search experiments, can further bound the values of $\delta$
that can be reached. This is the case for coherent interactions, for
which direct search experiments have already been able to set
stringent bounds. For $m_\chi=50$ GeV, the upper limit on
$\xi\sigma_c$ is shown in Fig. \ref{fig:coer_m50} by a vertical dashed
line. From this figure, we can therefore see that once the direct
detection constraint is taken into account, the largest values of
$\delta$ which can be obtained are of the order of
$10^{-6}$--$10^{-7}$, which are far too small to lead to observable
effects, as already discussed at the end of
Sect. \ref{sec:structure}. Once rescaling is considered, the
behaviours of the iso--$\delta$ and $N_\chi^{\rm max}$ lines plotted
in Fig. \ref{fig:coer_m50_rescal} show that $\delta$'s of the order of
$10^{-3}$--$10^{-2}$ can be reached. These values are close to the
necessary (but not sufficient) condition for WIMPs to have observable
effects on the Sun ({\em i.e. $\delta\gsim 10^{-2}$}), introduced in
Sect. \ref{sec:structure}, and therefore they need to be analyzed in
detail.

In the case of spin--dependent interactions, direct search experiments
are also setting bounds on $\xi\sigma_{sd}$, but these limits are not
model independent since they rely on the way the interaction of WIMPs
with the spin of the nucleus is realized \cite{bdmsbi}. For the
representative case of a Z--boson mediated interaction and for
$m_\chi=50$ GeV, the present upper limit is shown in
Fig. \ref{fig:spin_m50} by a dotted vertical line. Since this line
cannot be assumed to be a model independent bound, in the case of
spin--dependent interactions we are left with the possibility to reach
values of $\delta\sim10$.

Up to this point we have quantified the amount of possible deviations
induced by a generic WIMP in terms of the parameter $\delta$. We now
turn to discuss the possibility to experimentally observe such
modifications by means of helioseismology and of the Boron solar
neutrino flux.

\section{WIMPs and Helioseismology}
\label{sec:helioseismology}

\begin{figure} \centering
  \includegraphics[width=1.1\textwidth]{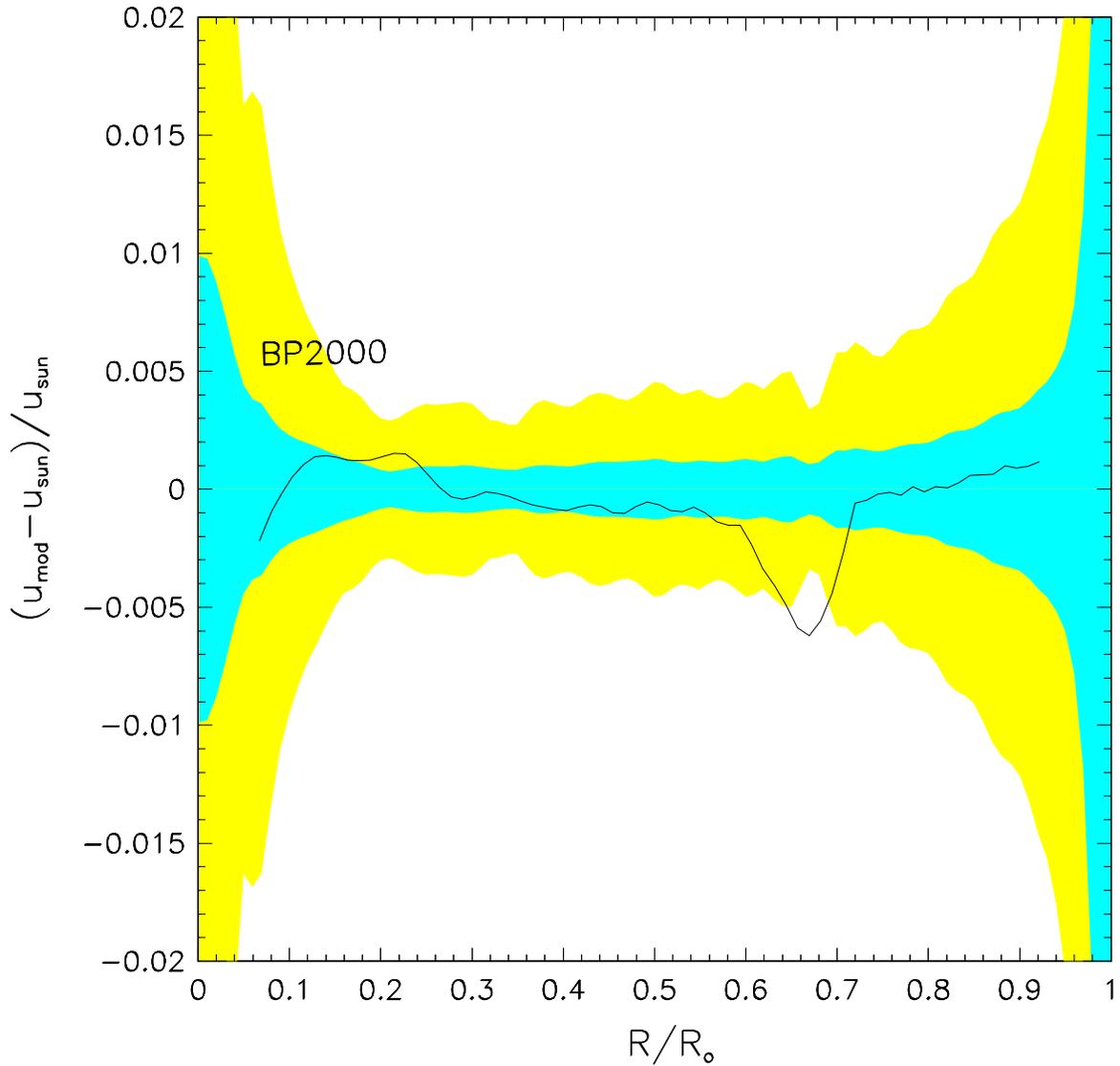}
\vspace{-50pt} \caption{Uncertainty bands in the reconstruction of the
squared isothermal sound speed $u$ in the Sun interior. The light
(yellow) region corresponds to the conventional ``$3\sigma$'' error
band, while the dark (blue) band identifies the conventional
``$1\sigma$'' range. The solid line corresponds to the fractional
difference between $u$ determinaed from observational data and $u$
calculated from the BP2000 solar model \cite{BP2000}.
\label{fig:speed0}}
\end{figure}

Helioseismology has provided very accurate information on the solar
structure and it has been able to establish severe constraints and
tests of the solar standard model (SSM) calculations. For instance,
helioseismology accurately determines the depth of the convective
envelope $R_b$ and the photospheric helium abundance
$Y_{ph}$. Moreover, by inversion of helioseismic data, one can
determine the (squared isothermal) sound speed in the solar interior,
$u= P/\rho$, with high accuracy ($P$ and $\rho$ denote the pressure
and density inside the Sun). In Fig. \ref{fig:speed0}, we show the
uncertainty in the helioseismic determination of $u$ as a function of
the radial coordinate $R/R_{\odot}$.  The light band corresponds to
the so-called ``$3\sigma$'' errors which have been estimated
conservatively by adding linearly all known individual uncertainties
\cite{Ricci:1997}.  If uncertainties are added in quadrature, the
global error is about one third. This yields the so called
``$1\sigma$'' errors which are shown by the dark band in
Fig.\ref{fig:speed0}. This latter procedure was also used by Bahcall
et al. \cite{Bahcall:speed} with similar results. For intermediate
values of the radial coordinate, the uncertainty in the squared sound
speed is at the level of $10^{-3}$, while in the innermost part of the
Sun, as well as on the surface, it is at the percent level. We remark
that, as found in Refs. \cite{Ricci:1997} and \cite{Bahcall:speed},
uncertainties corresponding to statistical errors of the frequency
measurements are generally much smaller than the ``systematical
uncertainties'' of the inversion method. These latter include the
choice of the reference model and of the free parameters in the
inversion procedure. These uncertainties are particularly important
close to the solar center, since there are very few {\it p-}modes that
sample this region well. The errors quoted above include both
statistical and systematic uncertainties. On the other hand, Refs.
\cite{turck1} and \cite{turck2} present sound speed profiles with
errors which look significantly smaller than our estimate.  This
occurs since ''the quoted uncertainties include only the contribution
arising from the frequency observations'', as clearly stated in
\cite{turck2}.

Fig. \ref{fig:speed0} also shows the difference between $u$ as
predicted by the SSM of Ref.~\cite{BP2000} and the helioseismological
determination $u_{sun}$, normalized to $u_{sun}$.  We notice that the
SSM accurately reproduces $u$ for all the radial profile.

\begin{figure} \centering
  \includegraphics[width=1.1\textwidth]{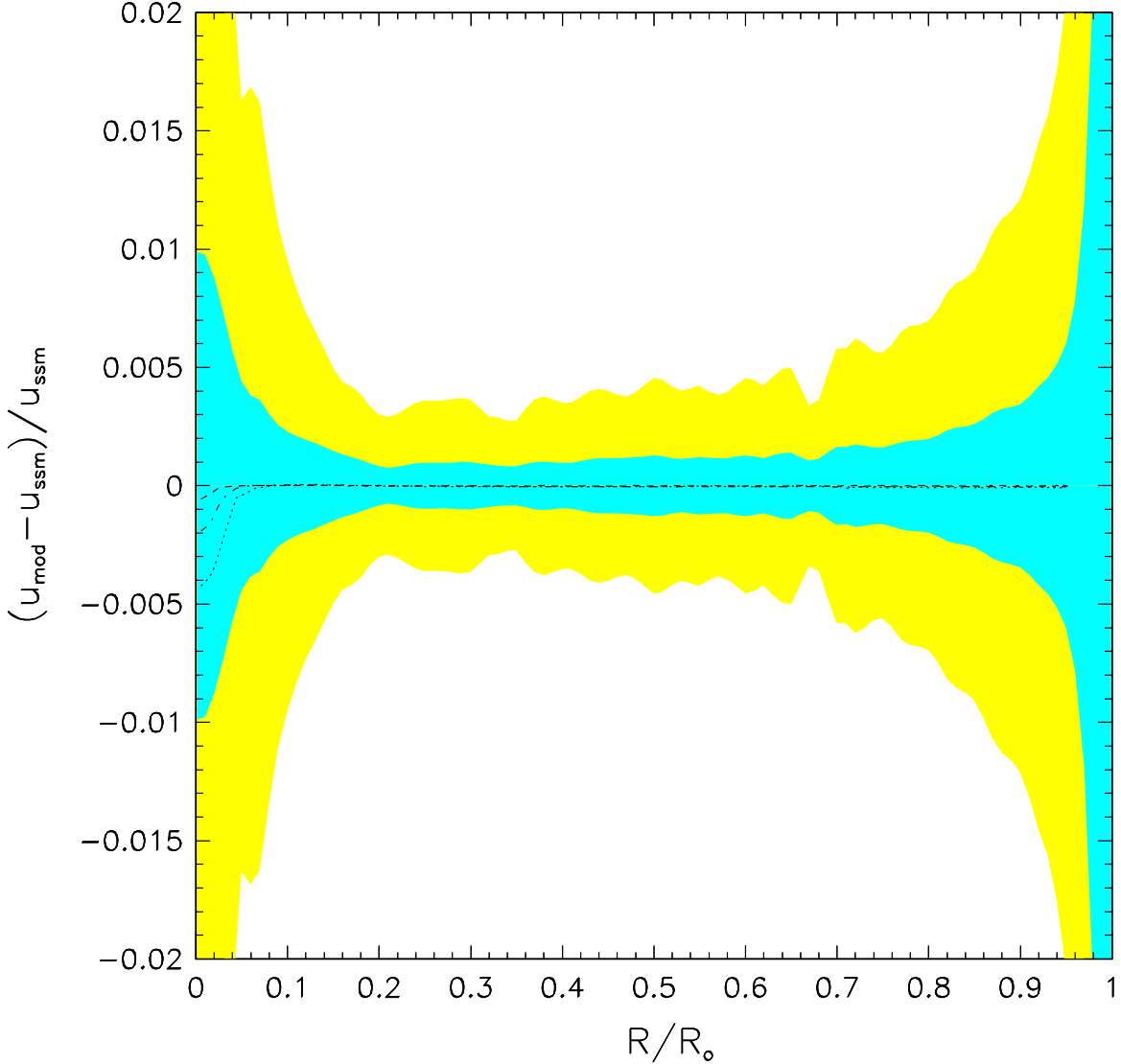}
\vspace{-50pt} \caption{Relative difference of the squared isothermal
sound speed in the Sun interior $u$ calculated for a standard solar
model and for solar models with accreting WIMPs, as a function of the
radial coordinate $R/R_{\odot}$. The light (yellow) region corresponds
to the conventional ``$3\sigma$'' error band, while the dark (blue)
band identifies the conventional ``$1\sigma$'' range. The dotted,
dot-dashed and dashed lines refer to solar models with accreting WIMPs
of masses $m_\chi = 30$, 50, 100 GeV and coherent scattering cross
sections which, for each mass, maximize the effect of the presence of
WIMPs in the Sun.
\label{fig:speed}}
\end{figure}

In order to understand whether helioseismology can tell us something
about WIMPs, we have constructed solar models for values of the WIMP
parameters which maximize the role of WIMPs in the energetics of the
Sun, irrespective of possible bounds derivable from direct search
experiments. As discussed in the previous Section, this corresponds to
$\delta \sim 10$. For $m_\chi=50$ GeV, this is shown, for instance, by
the large dot on the ``no-annihilation line'' in
Fig.\ref{fig:coer_m50}.
We have taken into account the time evolution of the number of WIMPs
in the Sun and we have described WIMP energy transport at each stage
of Sun evolution according to Eq. (\ref{general}). The sound speed
profiles obtained for the representative WIMP masses $m_{\chi}=30$, 50
and 100 GeV are compared with the sound speed profile of the SSM in
Fig.\ref{fig:speed}.  Larger WIMP masses lead to even smaller
effects. For any WIMP mass, smaller values of $\delta$ clearly
correspond to smaller effects.

The first observation is that, even in the most favourable situation,
WIMPs produce a modification of the sound speed profile which is much
smaller than the accuracy of present helioseismic determinations.  The
smallness of the effect is explained by two main reasons.  First,
WIMPs with masses larger than tens of GeV are confined in a small
region of the Sun of radius $R\lsim 0.02 R_{\odot}$.  In this region
the relevant physical quantities (temperature, luminosity, etc.) have
small variations. As a consequence, even if WIMPs affect sizeably the
gradient of these quantities ({\em e.g.} the temperature gradient),
this translates into small modifications of their ``global'' radial
profile. Second, a compensation occurs between two different effects.
This can be easily understood in terms of the fact that, from the
perfect gas law, which describes to a good approximation most of the
solar core, one has $u=P/\rho\simeq T/ \mu$, where $\mu$ is the mean
molecular weight. WIMPs produce a quasi-isothermal core, thus
decreasing the central temperature $T_c$ of the Sun.  At the same
time, however, with a lower $T_c$ one has smaller nuclear reaction
rates. This translates into a slower Helium production at the center
of the Sun and, thus, into a decrease of the mean molecular weight.
When considering $u\simeq T/ \mu$, the effects of the variations of
$T$ and $\mu$ compensate, leaving the sound speed profile almost
unaffected.

The results shown in Fig. \ref{fig:speed} refer to a situation which
maximizes the effect of the presence of WIMPs in the Sun: $\delta \sim
10$. As discussed in the previous Section, for spin--dependent
interactions, this value of $\delta$ can be reached, while for
coherent interaction $\delta$'s of the order of at most
$10^{-3}$--$10^{-2}$ can be obtained. Our results show that in neither
case (spin--depenent nor coherent) the presence of WIMPs in the Sun
can affect $u$ to a level which is currently accessible by
helioseismology.

We notice also that the necessary (but not sufficient) condition
stated in Sect. \ref{sec:structure}, {\em i.e.} that $\delta$ must be
larger that about $10^{-2}$ in order to possibly have an observable
effect, may be reinforced: in fact our detailed calculation shows
that, for WIMP masses above 30 GeV, not even $\delta\sim10$ is
sufficient to produce a currently observable effect.

In summary, the results of this Section show that no significant
information about WIMPs with masses $m_{\chi} \gsim 30$ GeV can be
obtained at present from helioseismic data, for any values of the
WIMP-nucleon cross section.

Conclusions conflicting with ours were derived in
Ref. \cite{Lopes:2001ra}.  This may be traced back to the fact that in
Ref. \cite{Lopes:2001ra} the accuracy of the squared sound speed is
taken to be of about 0.1\% also in the innermost part of the Sun core
relevant for WIMP confinement, say $R\lsim 0.02 R_{\odot}$.
We argued above that, at these small radii, the accuracy of the 
helioseismic determination is significantly worse.

\section{WIMPs and the Boron neutrino flux}
\label{boron}

From the previous discussion, one could expect that the Sun central
temperature $T_c$ is more sensitive than the central sound speed to
the presence of WIMPs in the Sun. This is actually the case, as can be
seen in Table \ref{tab:variations}, which shows that WIMPs produce
variations of $T_c$ at the percent level, while the central sound
speed $u_c$ is affected at most at the level of $10^{-3}$.

Unfortunately, the central temperature of the Sun is not directly
observable. The most direct information on it comes from the the
measurements of the solar neutrino fluxes. In particular, the
comparison between the SNO charged current and Super-Kamiokande data
\cite{Ahmad:2001an,Fukuda:2001nj,Villante:1998pe,Fiorentini:2001jt,
Fogli:2001vr,Fogli:2002ts} has allowed to determine, in a model
independent way, the $^{8}{\rm B}$ neutrino flux produced in the Sun
with an accuracy of about 20\%
\footnote{The SNO experiment has recently reported the observations of
neutral current $\nu$ interactions on deuterium \cite{Ahmad:2002jz}
confirming the $^{8}{\rm B}$ neutrino flux predicted by SSM with an
accuracy of about 12\%. This result is obtained under the assumption
that the solar neutrino energy spectrum is undistorted.  For this
reason, in our analysis, we refer to the result obtained by comparing
SNO charged current and Super-Kamiokande data, which instead do not
rely on this assumption, see
Refs. \cite{Villante:1998pe,Fogli:2001nn}.}.  This result, {\it under
the assumption that the $^{8}{\rm B}$ neutrino flux scales with the
Sun central temperature as $\Phi_{\rm B} \propto T_{c}^{20}$}, can be
used to determine $T_{c}$ at the 1\% level \cite{Fiorentini:2001et}.

\begin{figure} \centering
  \rotatebox{90}{\includegraphics[width=0.70\textwidth]{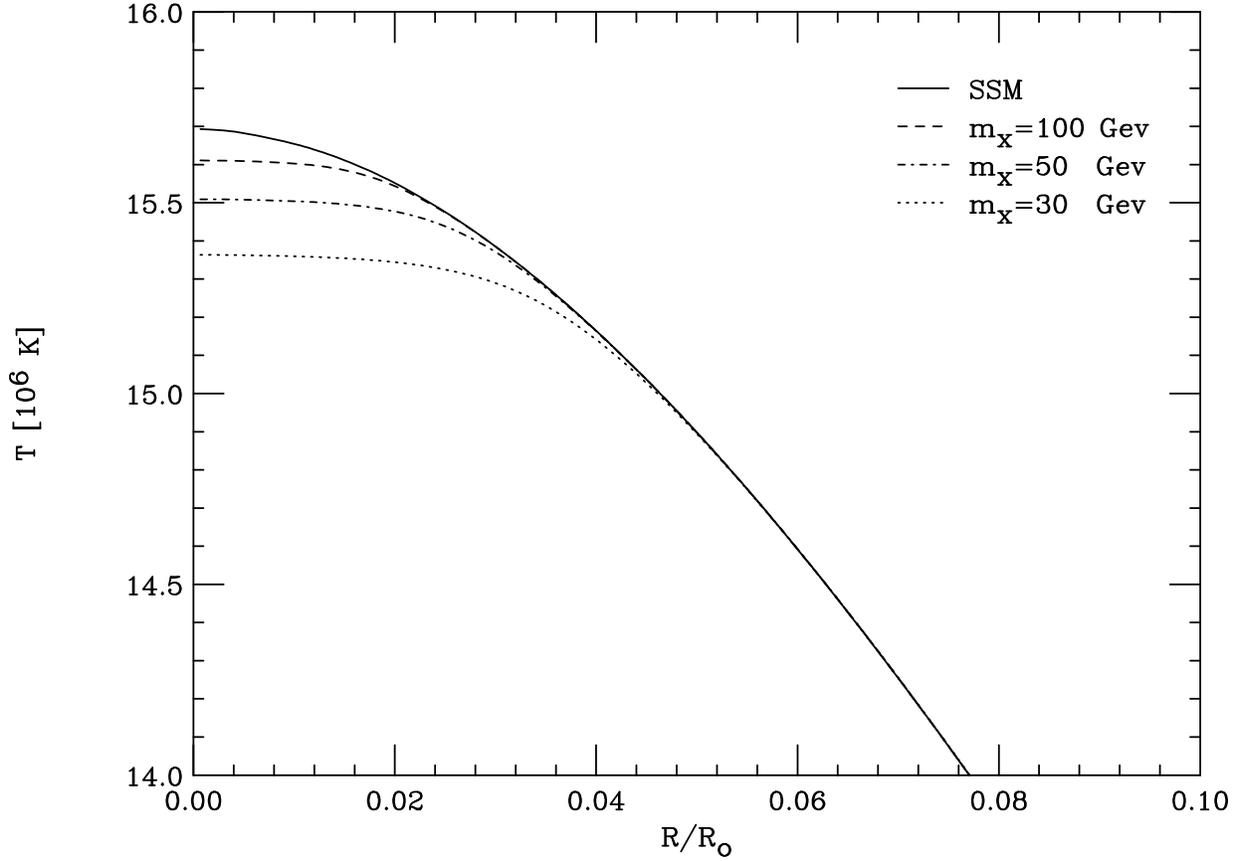}}
  \caption{Temperature profile of the standard solar
  model and of models of the Sun where WIMPs with masses of 30, 50 and
  100 GeV are present. The coherent WIMP--nucleus scattering cross
  sections are chosen, for each mass, in order to maximize the effect
  of the presence of WIMPs in the Sun.
\label{fig:temperature}}
\end{figure}

However, the presence of WIMPs in the core of the Sun produces a very
peculiar modification of the temperature profile for which the scaling
law $\Phi_{\rm B} \propto T_{c}^{20}$ is not valid. WIMPs can sizeably
affect the temperature profile only in the innermost part of the Sun,
and in particular in a region which is smaller than the Boron neutrino
production region ($r_B \simeq
0.05 R_{\odot}$). Fig.\ref{fig:temperature} shows the radial profile
of the Sun temperature in the SSM and in models of the Sun where WIMPs
with masses of 30, 50 and 100 GeV are present. Again, we have chosen
the WIMP parameters which maximize the effect of their presence in the
Sun. Fig.\ref{fig:temperature} clearly shows that even though the
central temperature of the Sun can be lowered by the presence of WIMPs
at the percent level (as it is reported also in Table
\ref{tab:variations}), nevertheless the temperature at $r_B \simeq
0.05 R_{\odot}$, where the Boron neutrino production is maximal, is
left practically unchanged.

In order to perform a quantitative analysis, one has therefore to look
directly at the Boron neutrino flux and not simply at the central
temperature $T_c$. If this is done, it turns out that the solar models
of Fig. \ref{fig:temperature}, which correspond to WIMP masses equal
to 30, 50 and 100 GeV, predict variations of the Boron neutrino flux
equal to: $\delta \Phi_{\rm B}/ \Phi_{\rm B}\simeq -3.6 \%$, $-1.0\%$
and $-0.1\%$, respectively. These variations are well below the
accuracy of the present determination of the Boron neutrino flux. We
remind that, for each WIMP mass, we have optimized the other WIMP
parameters in order to maximize the effect. Larger masses and WIMP
annihilation would lead to even smaller deviations.

In conclusion, our results allow us to conclude that no information
about WIMPs with massess $m_{\chi} \gsim 30$ GeV can be obtained from our
present knowledge of the Boron neutrino flux. 

A previous independent analysis \cite{Lopes:2001ig} concluded that
WIMPs lighter than 60 GeV are in conflict with the present accuracy in
the determination of the Boron neutrino flux, based on the fact that
the presence of WIMPs in the core of the Sun can modify the central
temperature $T_c$ at the 1\% level. We should stress that this lower
limit on the WIMP mass does not apply, since, as we have discussed
above, WIMPs, even though capable of changing $T_c$ at the percent
level, nevertheless are too much concentrated in the interior of the
Sun to produce sizeable modifications of the Boron neutrino flux,
which is the quantity experimentally measured.

\section{One realization of WIMP: the neutralino}
\label{sec:neutralino}

\begin{figure} \centering
  \includegraphics[width=1.1\textwidth]{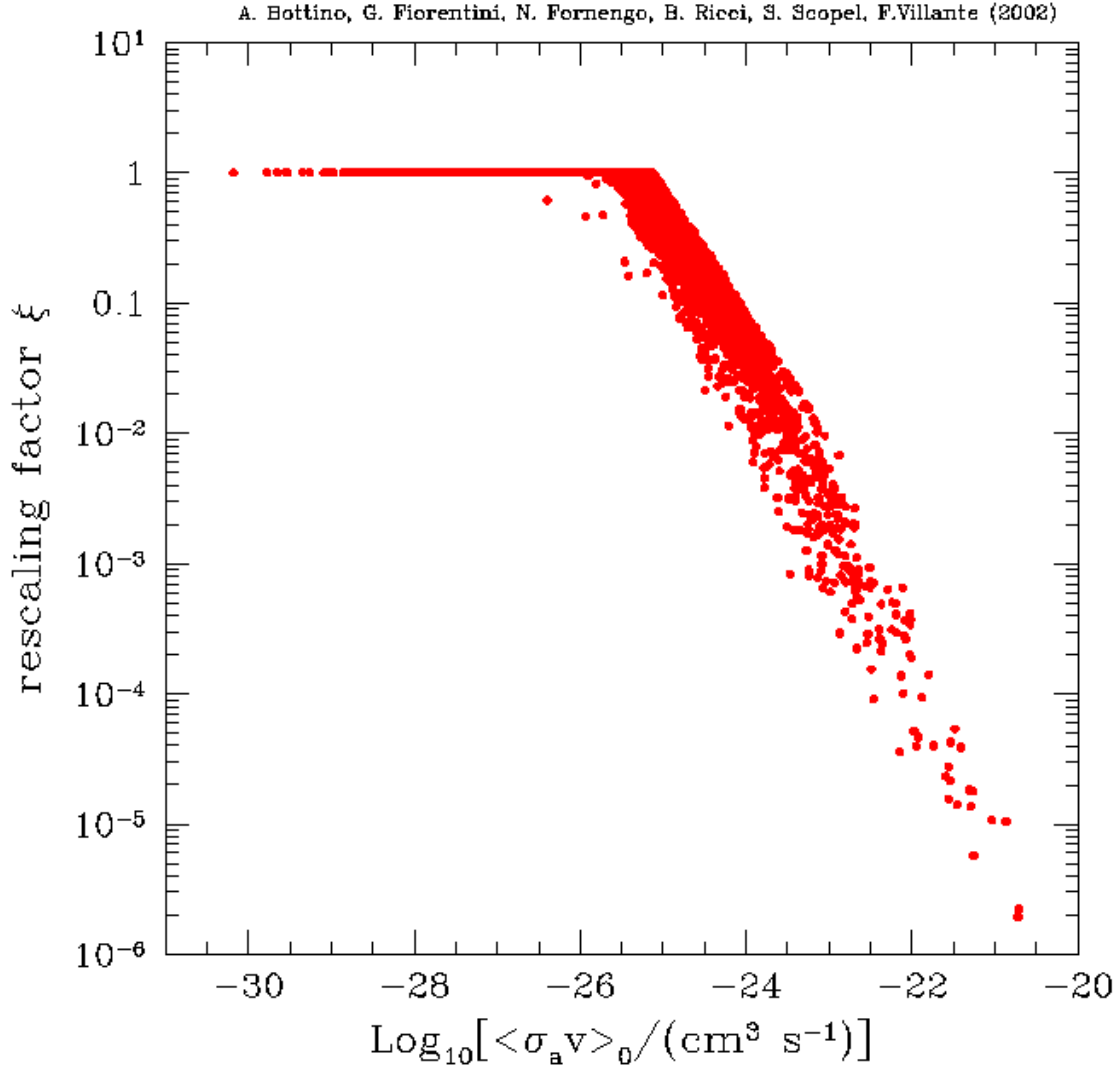}
\vspace{-50pt} \caption{Rescaling parameter $\xi = \min(1, \Omega_\chi
h^2/0.05)$ as a function of the zero--temperature thermally--averaged
neutralino self--annihilation cross section $\langle \sigma_a v
\rangle_0$ in the effective minimal supersymmetric standard model
(MSSM).
\label{fig:mssm_rescaling}}
\end{figure}

\begin{figure} \centering
\includegraphics[width=1.1\textwidth]{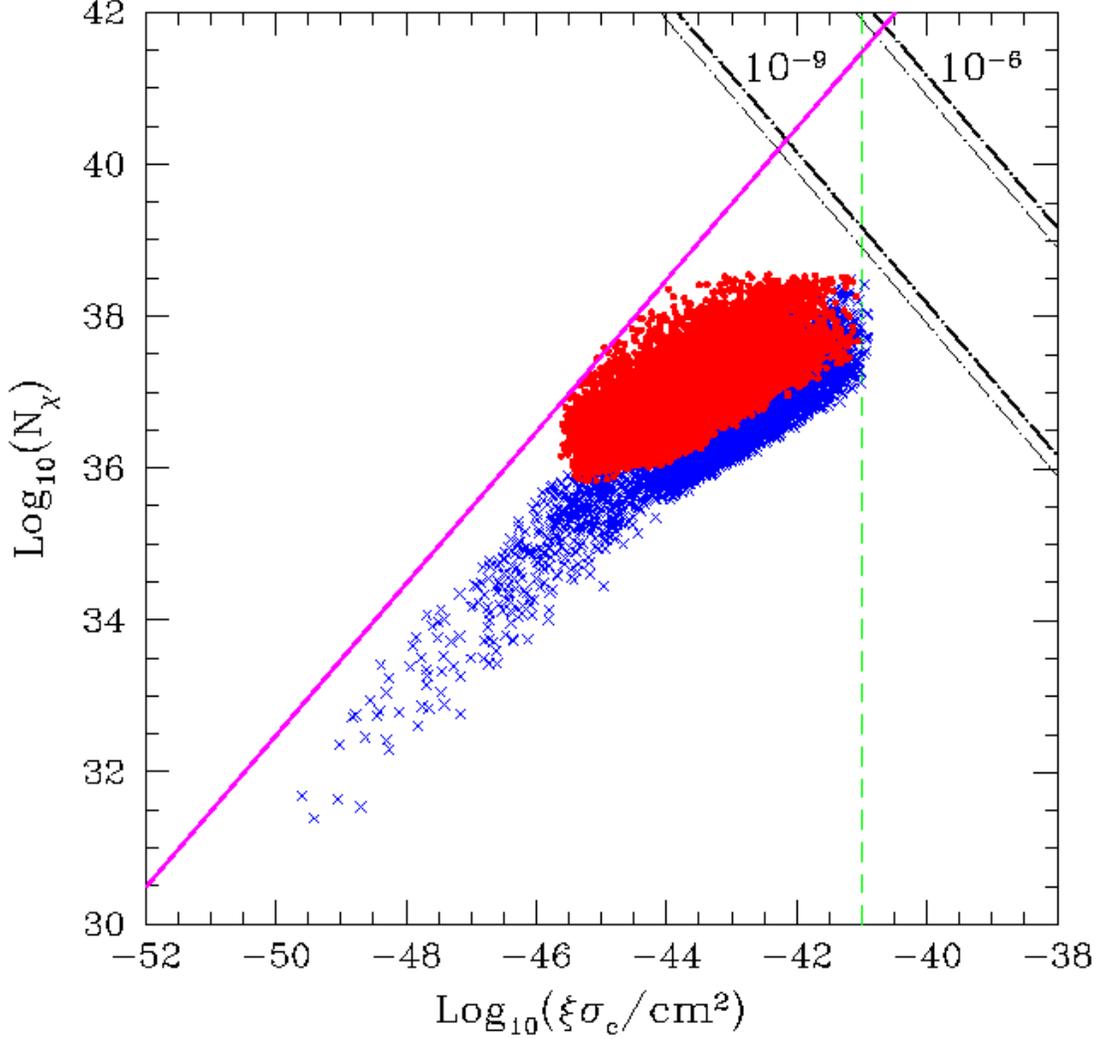}
\vspace{-50pt}
\caption{Total number $N_\chi$ of neutralinos accumulated in the Sun
as a function of $\xi \sigma_c$ in the effective MSSM. Only
configurations where the total capture rate is dominated by coherent
interactions are displayed. Neutralino masses are in the range
$50~{\rm GeV} < m_\chi < 500~{\rm GeV}$. (Red) dots denote
configurations where the neutralino is the dominant dark matter
component: $0.05 < \Omega_\chi h^2 < 0.3$ ({\em i.e.}: $\xi =
1$). (Blue) crosses refer to configurations where the neutralino is a
sub--dominant dark matter component: $\Omega_\chi h^2 < 0.05$ ({\em
i.e.}: $\xi < 1$).  The slanted (pink) solid line shows $N_\chi$ for
$m_\chi=50$ GeV and for the limiting case of non-annihilating
WIMPs. The (black) dot--dashed lines denote iso--$\delta$
contours. For each pair, the upper line refers to the iso--$\delta$
curve for $m_\chi=50$ GeV and $\xi=1$, while the lower line refers to
the same iso--$\delta$ and $\xi$, for $m_\chi=500$ GeV. The vertical
(green) dashed lines indicate the upper limit on $\xi \sigma_p$ for
coherent interactions.
\label{fig:mssm_number_coer}}
\end{figure}

\begin{figure} \centering
  \includegraphics[width=1.1\textwidth]{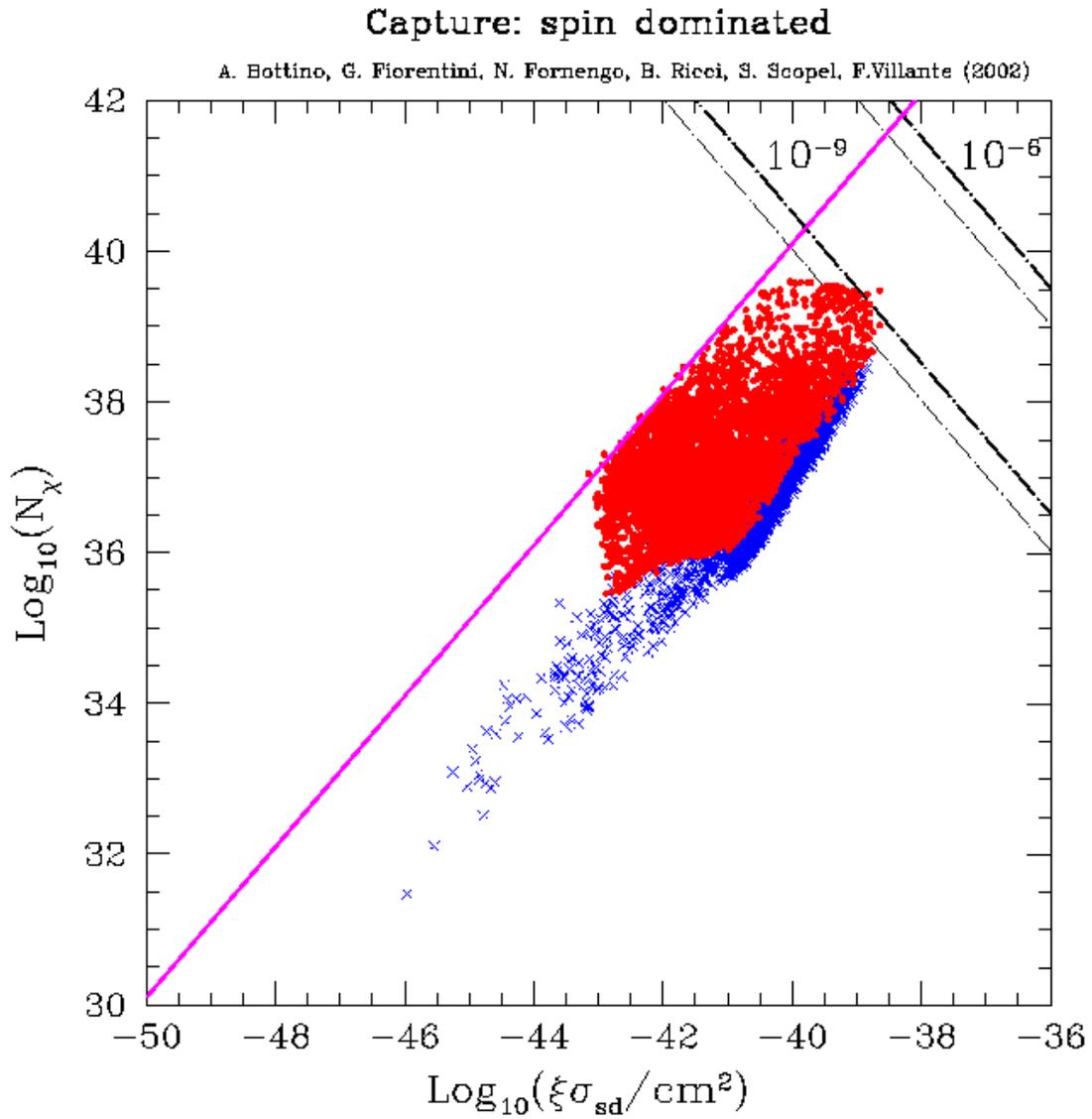}
\vspace{-50pt} 
\caption{The same as in
Fig. \ref{fig:mssm_number_coer}, for MSSM configurations where the
total capture rate is dominated by spin--dependent interactions.
\label{fig:mssm_number_spin}}
\end{figure}

It is instructive to discuss how some of the previous properties are
realized in the case of one of the favourite WIMP candidates: the
neutralino. In
Figs. \ref{fig:mssm_rescaling}--\ref{fig:mssm_number_spin} we show
some of the relevant properties.

The supersymmetric scheme we are adopting here is an effective Minimal
Supersymmetric extension of the Standard Model (MSSM). Its parameters
are defined directly at the electroweak scale and represent the
minimum set necessary to shape the essentials of the theoretical
structure of the model and of its particle content. We refer to Ref.
\cite{mssm_details} for details on the theoretical aspects and on the
updated experimental bounds.

The neutralino--nucleon cross sections, both coherent and
spin--dependent, are calculated in terms of the couplings and
supersymmetric particle masses as discussed, for instance, in
Refs. \cite{bdmsbi,noi}. The parameter $\xi$ is evaluated by using a
standard rescaling procedure \cite{gst} which relates the fraction of
local neutralino dark matter in the Galaxy to its fractional abundance
in the Universe, {\it i.e.} to its relic abundance $\Omega_{\chi}
h^2$:
\begin{equation}
\xi = \min \left [1,
\frac{\Omega_{\chi} h^2}{(\Omega_m h^2)_{\rm min}} \right ] \, , 
\label{eq:rescal}
\end{equation}
where $(\Omega_m h^2)_{\rm min}$ is the minimum value of $\Omega_m h^2$
compatible with halo properties. Here we set $(\Omega_m
h^2)_{\rm min}=0.05$. The present--day neutralino relic abundance has
been calculated according to Ref.\cite{omega}.

To establish some relevant order of magnitude for the annihilation
cross--section $\sigma_a$ it is useful to recall the approximate
expression: $\Omega_{\chi} h^2 \sim {3 \cdot 10^{-27} {\rm cm}^3~ {\rm
s}^{-1}}/{\langle \sigma_{a} \; v \rangle_{int}}$, where $\langle
\sigma_{a} \; v \rangle_{int}$ denotes the thermal average of the
product $(\sigma_{a} \cdot v)$ integrated from the freeze--out
temperature $T_f$ to the present--day one.  For instance the upper
bound $\Omega_{\chi} h^2 \lsim 0.3$ translates into a lower bound
$\langle \sigma_{a} \; v \rangle_{int}\, \gsim 10^{-26}$ cm$^3$
sec$^{-1}$.

In Fig. \ref{fig:mssm_rescaling} we display the correlation between
$\xi$ and $\langle \sigma_{a} \; v \rangle_{0}$, calculated by varying
the MSSM parameters and applying to the model the relevant
experimental constraints. Fig. \ref{fig:mssm_rescaling} shows that for
a neutralino in the MSSM, the rescaling factor is always smaller than
1 for values of $\langle \sigma_{a} \; v \rangle_{0}$ larger than
about $10^{-27}$--$10^{-26}$ cm$^3$ s$^{-1}$. Notice that $\langle
\sigma_{a} \; v \rangle_{0}$ denotes the thermal average of the
product $(\sigma_{a} \cdot v)$ in the Sun's core, and does not
coincide with the relevant thermal average which enters in the
calculation of the relic abundance.

We turn now to the graphic representations in the plane $\xi
\sigma_p$--$N_{\chi}$. Fig. \ref{fig:mssm_number_coer}
(Fig. \ref{fig:mssm_number_spin}) shows the scatter plot of the MSSM
configurations for which the total capture rate is dominated by
coherent (spin--dependent) interactions and for the neutralino mass
range: $50~{\rm GeV} < m_\chi < 500~{\rm GeV}$, the lower bound
arising from experimental constraints on the MSSM. The points are
labelled according to the values of the neutralino relic abundance:
dots denote configurations where the neutralino is the dominant dark
matter component: $0.05 < \Omega_\chi h^2 < 0.3$, {\it i.e.}  $\xi =
1$; crosses refer to configurations where the neutralino is a
sub--dominant dark matter component: $\Omega_\chi h^2 < 0.05$, {\it
i.e.}  $\xi < 1$. The slanted solid line shows $N_\chi$ for
$m_\chi=50$ GeV and for the limiting case of non-annihilating
WIMPs. This line represents, for each of the two limiting cases of
coherent and spin-dependent domination, the actual upper limit for
$N_\chi$ for neutralinos in the effective MSSM, due to the lower bound
on the neutralino mass quoted above. The dot--dashed lines denote
iso--$\delta$ contours for two representative values of $\delta$. For
each pair, the upper line refers to $m_\chi=50$ GeV and $\xi=1$, while
the lower line refers to $m_\chi=500$ GeV and $\xi=1$.

By comparing the scatter plot with the iso--$\delta$ lines, we can
notice that the configurations with no rescaling ($\xi=1$) are
compatible with values of $\delta$ smaller than about $10^{-9}$, both
for coherent and spin--dependent interactions. We remind that these
are configurations with large neutralino relic abundance.  In the case
of configurations with $\xi<1$, we have to consider the shift of the
iso--$\delta$ lines shown in Fig. \ref{fig:coer_m50_rescal} and the
range of values for $\xi$ in the MSSM shown in
Fig. \ref{fig:mssm_rescaling}. In this case, we obtain that the
maximal effect induced by subdominant relic neutralino is for values
of $\delta$ of the order of $10^{-6}$ for the coherent case and
$10^{-4}$ for the spin--dependent one.

We can therefore conclude that the maximal effects induced in the Sun
by relic neutralinos are much smaller than those valid for a generic
WIMP and discussed in Sect. \ref{sec:number}. This implies that the
specific case of the neutralino has even smaller chances to be bounded
by the study of solar properties.

\section{Conclusions} 
\label{sec:finale}

Our understanding of solar physics has significantly advanced  
in recent times, because of remarkable improvements in helioseismology 
and of new data on solar neutrinos. In the present paper we have 
addressed the question of whether these new developments put solar 
physics in a position to provide  constraints on the 
possible presence of WIMPs in the core of the Sun, for values of 
WIMP parameters under current exploration in WIMP searches. 
We summarize here our main results:
\begin{itemize}
\item
We have provided a quantitative criterium to determine whether a
putative WIMP candidate could produce observable modifications of the
solar structure. Namely, we have introduced the parameter $\delta$,
defined as the ratio between the WIMP luminosity and the radiative
luminosity at the center of the Sun (see Eq.(\ref{eq:delta})), which
can be analytically calculated as a function of the WIMP-nucleon
scattering cross section $\sigma_p$ and the number of WIMPs in the Sun
$N_{\chi}$.  By considering the uncertainty in the radiative opacity,
one finds that $\delta \gsim 10^{-2}$ is a necessary (but not
sufficient) condition for WIMPs to have observable effects on the Sun.
\item
We have calculated the number of WIMPs in the Sun for a generic WIMP
candidate as a function of the WIMP-nucleon scattering cross section
$\sigma_p$ and of the WIMP-antiWIMP pair annihilation cross-section
$\sigma_a$, both for coherent and spin-dependent WIMP interactions. We
have shown that a value $\delta \sim 10$ can be reached, if
WIMP-antiWIMP annihilation is negligible and $\sigma_p \sim 10^{-37}
\;{\rm cm}^2$ (coherent scattering) or $\sigma_p \sim 10^{-34} \;{\rm
cm}^2$ (spin-dependent interactions).  For coherent WIMP scattering,
such large values for the cross sections are excluded by direct search
experiments, unless WIMPs give a subdominant contribution to dark
matter of our galaxy.  In this case, if one takes into account direct
search bounds ($\xi\sigma_p\lsim 10^{-41} \;{\rm cm}^2 $, where $\xi$
is the rescaling parameter which accounts for the actual fraction of
local dark matter to be ascribed to the given WIMP), one obtains
$\delta\lsim 10^{-6}$ in the assumption of no rescaling ($\xi =1$) and
$\delta\lsim 10^{-3}$ in the general case ($\xi \leq 1$).
\item
We have considered the neutralino as a specific WIMP candidate. We
have calculated the number of neutralinos in the Sun for MSSM
configurations compatible with present experimental and cosmological
constraints.  For each configuration, we determined the rescaling
parameter $\xi$ by the standard rescaling procedure described in
Eq.~(\ref{eq:rescal}).  In the case of no rescaling ($\xi=1$), we
obtained $\delta\lsim 10^{-9}$. For subdominant neutralinos ($\xi\leq
1$), we obtained $\delta\lsim 10^{-6}$ for configurations in which the
capture rate is dominated by coherent scattering and $\delta\lsim
10^{-4}$ for configurations in which the dominant contribution is due
to spin dependent interactions.
\item
The previous points already show that solar physics is not competitive
with direct experiments in a large part of the WIMP parameter space
under current exploration in WIMP searches.  In order to complete our
analysis and to understand whether some information could be obtained
from the present helioseismic and solar neutrino data, we have
constructed solar models, for WIMP masses above 30 GeV, choosing the
value of the WIMP parameters which maximize the effect of WIMPs on the
Sun ($\delta \sim 10$).  As a result of the presence of WIMPs in the
Sun, we obtained variations of the sound speed profile and of the
$^8$B neutrino flux which are within the current experimental
uncertainties.  The smallness of the effects is essentially due to the
the smallness of the WIMP estension region which, for WIMPs with
masses larger than tens of GeV, is $r_\chi \leq 0.02 R_{\odot}$.
\end{itemize}

In conclusion, no constraints can be derived at present from solar
physics for WIMPs with masses above 30 GeV. Our conclusions are at
variance with results derived in Refs.
(\cite{Lopes:2001ra,Lopes:2001ig}). The origins of these disagreements
have been elucidated in the present paper.

\appendix  
\section{}

We give here for completeness the expression of the factor $S_i$,
introduced in Eq.(\ref{eq:capture}), as derived from
Ref.\cite{Gould:1987}. It is given by:
\be
S_i=\frac{\langle F_i \rangle_i}{\langle \phi \rangle_i~\xi(\infty)} ,
\ee
where the brackets indicate an average over the mass density profile
$\rho_i(r)$ of the $i$--th nucleus in the Sun:
\begin{equation}
\langle f \rangle_i \equiv \frac{1}{M_i}\int_0^{R_\odot} 4\pi r^2
dr\; \rho_i(r)\; f(r)
\label{eq:mass_ave}
\end{equation}
All the quantities are defined in Section III with the exception of
$F_i$, which is given by:
\bea
F_i &=& \frac{\bar{v}^2}{v_{esc}^2} \frac{1}{3 b \eta} 
\left\{ 
\left[\chi (-\hat{\eta},\hat{\eta})-\chi (\hat{A}_{-},\hat{A}_{+})
\right]
\frac{\exp(-a \hat{\eta}^2)}{(1+a)^{1/2}}- \right. \nonumber \\
&&\left. \left[\chi (-\check{\eta},\check{\eta})-\chi (\check{A}_{-},\check{A}_{+})
\right] \frac{\exp(-b \check{\eta}^2)}{(1+b)^{1/2}}\exp[-(a-b)A^2]
\right\}
\eea
The definition of the quantities $a$, $b$, $\eta$, $A$,
$\hat{A}_{+,-}$, $\check{A}_{+,-}$, $\hat{\eta}$,  and $\check{\eta}$
may be found in Ref.\cite{Gould:1987}.


\clearpage
\begin{table}
\caption{\label{tab:sun} Sun composition, average mass fractions of
the chemical elements in the Sun as obtained from
Ref.\cite{Bahcall:1995} and values of the average reduced gravitational
potential $\langle \phi \rangle_i$.}
\begin{tabular}{|c|c|c|}
\hline
Element &  $X_i$  & $\langle \phi \rangle_i$ \\
\hline
H    &   ~~0.71~~            &   ~~3.16~~ \\        
He   &   ~~0.27~~            &   ~~3.40~~ \\        
C    &   ~~0.30 $10^{-2}$~~  &   ~~3.23~~ \\
N    &   ~~0.93 $10^{-3}$~~  &   ~~3.23~~ \\
O    &   ~~0.84 $10^{-2}$~~  &   ~~3.23~~ \\
Ne   &   ~~0.17 $10^{-2}$~~  &   ~~3.23~~ \\
Na   &   ~~0.35 $10^{-4}$~~  &   ~~3.23~~ \\
Mg   &   ~~0.65 $10^{-3}$~~  &   ~~3.23~~ \\
Al   &   ~~0.56 $10^{-4}$~~  &   ~~3.23~~ \\
Si   &   ~~0.70 $10^{-3}$~~  &   ~~3.23~~ \\
P    &   ~~0.62 $10^{-5}$~~  &   ~~3.23~~ \\
S    &   ~~0.37 $10^{-3}$~~  &   ~~3.23~~ \\
Cl   &   ~~0.78 $10^{-5}$~~  &   ~~3.23~~ \\
Ar   &   ~~0.94 $10^{-4}$~~  &   ~~3.23~~ \\
Ca   &   ~~0.65 $10^{-4}$~~  &   ~~3.23~~ \\
Ti   &   ~~0.36 $10^{-5}$~~  &   ~~3.23~~ \\
Cr   &   ~~0.17 $10^{-4}$~~  &   ~~3.23~~ \\
Mn   &   ~~0.96 $10^{-5}$~~  &   ~~3.23~~ \\
Fe   &   ~~0.12 $10^{-2}$~~  &   ~~3.23~~ \\
Ni   &   ~~0.74 $10^{-4}$~~  &   ~~3.23~~ \\  
\hline
\end{tabular}
\end{table}

\clearpage
\begin{table}
\caption{\label{tab:variations} Fractional variations of the central
temperature of the Sun $T_c$, of the squared isothermal sound speed in
the Sun interior and of the $^8$B neutrino flux, calculated in solar
models where WIMPs are accreted in the Solar core. For each model, the
WIMP--nucleon scalar cross section has been set at the value which
maximizes the effect of the presence of WIMPs in the Sun.}
\begin{ruledtabular}
\begin{tabular}{|c|c|c|c|}
$m_{\chi}$ (GeV) & $\delta T_{c} / T_{c}$ & $\delta u_{c} / u_{c}$  & $\delta \Phi_{B} / \Phi_{B}$\\
\hline
30 & $-2.1\cdot10^{-2}$& $-4\cdot10^{-3}$ & $-4.5\cdot10^{-2}$\\
50 & $-1.2\cdot10^{-2}$& $-2\cdot10^{-3}$ & $-1.3\cdot10^{-2}$\\
100 & $-5.2\cdot10^{-3}$& $-6\cdot10^{-4}$& $-2.1\cdot10^{-3}$\\
\end{tabular}
\end{ruledtabular}
\end{table}


\clearpage

\clearpage

\clearpage

\clearpage

\clearpage

\clearpage

\clearpage

\clearpage

\clearpage

\end{document}